\documentclass[11pt,a4paper]{article}

\usepackage[margin=2.3cm, top=2.5cm, bottom=2.5cm]{geometry}

\usepackage{xcolor}
\usepackage{graphicx}
\usepackage{tikz}
\usepackage{tcolorbox}

\usepackage[T1]{fontenc}
\usepackage{lmodern} 
\usepackage{microtype}
\usepackage{amsfonts}

\usepackage{hyperref}
\usepackage{caption}
\usepackage{titlesec}
\usepackage{enumitem}

\usepackage{graphicx}
\usepackage{booktabs}
\usepackage{hyperref}
\usepackage{xspace}
\newcommand{\ie}{i.e.\xspace}
\usepackage{amsmath}
\usepackage[dvipsnames,table]{xcolor}
\usepackage{colortbl}
\usepackage[normalem]{ulem}
\newcommand{\mila}[1]{{\color{black}{#1}}}
\usepackage{multirow}
\usepackage{orcidlink}

\hypersetup{
    colorlinks=true,
    linkcolor=magenta!80!gray,
    urlcolor=magenta!80!gray,
    citecolor=magenta!80!gray
}

\titleformat{\section}
  {\sffamily\bfseries\Large\color{magenta!80!gray}}
  {\thesection}{0.8em}{}

\titleformat{\subsection}
  {\sffamily\bfseries\large\color{magenta!80!gray}}
  {\thesubsection}{0.8em}{}

\titleformat{\subsection}
  {\sffamily\bfseries\color{magenta!80!gray}}
  {\thesubsubsection}{0.8em}{}

\setlist[itemize]{leftmargin=1.5em, itemsep=0.4em, topsep=0.3em}

\begin{document}

\begin{center}
\vspace*{-0.5cm}

\vspace{0.8em}
{\color{magenta!80!gray}\hrule height 0.6pt}
\vspace{1.2em}

{\sffamily\bfseries\huge Frozen CLIP Priors for Robust \\Self-Supervised Poisson Inverse Problems \par}

\vspace{1.2em}
{\color{magenta!80!gray}\hrule height 0.6pt}
\vspace{1.5em}

{\sffamily\bfseries\large Laura C. Diaz-Delgado \orcidlink{0009-0003-9917-9505},
Emmanuel Martinez \orcidlink{0000-0002-6458-4258},
Henry Arguello \orcidlink{0000-0002-2202-253X} \par}
\vspace{0.4em}
{\sffamily\small Department of Computer Science, Universidad Industrial de Santander, Colombia\\\href{henarfu@uis.edu.co}{\color{magenta!80!gray}henarfu@uis.edu.co}\par}

\end{center}

\vspace{1em}

\begin{tcolorbox}[
    colframe=magenta!80!gray,
    colback=white,
    arc=10pt,
    boxrule=0.8pt,
    left=14pt, right=14pt, top=12pt, bottom=10pt
]
\begin{center}
    {\sffamily\bfseries\large\color{magenta!80!gray} Abstract}
\end{center}
\vspace{-0.3em}
\noindent
Self-supervised learning for imaging inverse problems is increasingly important in photon-limited settings, where acquiring clean ground truth is impractical and reconstruction must remain stable under dataset and acquisition shifts. This challenge is amplified under Poisson noise, whose signal-dependent statistics interact with sampling operators (e.g., CFA mosaicing). Meanwhile, foundation vision encoders trained at web scale offer distortion-invariant, content-related representations that generalize well across domains, suggesting a promising route to build priors that transfer beyond the training distribution without expensive fine-tuning. This paper proposes an ADMM-inspired unrolled plug-and-play solver for Poisson inverse problems that decouples a closed-form data-consistency update from a parameter-efficient prior. The prior is implemented as a lightweight decoder operating on frozen CLIP RN50 dense multi-scale features, adapting foundation representations with less trainable parameters. For self-supervision, the method integrates GR2R measurement-domain re-corruption with an Equivariant Imaging regularizer via virtual acquisitions. Experiments on Poisson CFA demosaicing and deblurring show competitive quality, improved robustness under shifts, and self-supervised performance approaching supervised training.
\vspace{0.4em}
{\color{magenta!40}\hrule height 0.4pt}
\vspace{0.3em}
\vspace{0.8em}
\small\sffamily
\noindent\textbf{Official Page:} \url{https://github.com/LauraCD2/frozen-clip-priors} \\[2pt]

\end{tcolorbox}

\vspace{1.5em}

\section{Introduction}
\label{sec:intro}

Imaging inverse problems, including deblurring, demosaicing, super-resolution, inpainting, and compressed sensing, are central to computer vision and computational imaging, as many acquisition pipelines produce indirect, incomplete~\cite{tachella2022unsupervised}, or physically constrained measurements that must be inverted to recover \mila{image} content~\cite{aggarwal2018modl,venkatakrishnan2013plug}. Such problems are typically ill-posed: a single observation can correspond to multiple plausible reconstructions, so successful recovery hinges on strong priors or regularization~\cite{demmel1987condition,fernandez2020curse,romano2017red}. In many high-impact settings (e.g., microscopy~\cite{zhang2019poisson}, medical imaging~\cite{rodrigues2008denoising}, and low-light photography~\cite{chen2018learning}), collecting clean ground truth is expensive or impossible, making self-supervised reconstruction a requirement in practice rather than a  preference~\cite{tachella2022unsupervised,chen2021equivariant}. This need is crucial in photon-limited regimes, where measurements are degraded by signal-dependent Poisson noise~\cite{hohage2016inverse}. In such settings, the noise changes with the signal level, interacts with sampling operators such as Color Filter Array (CFA) mosaicing, and can destabilize objectives and architectures designed under Gaussian assumptions~\cite{hohage2016inverse,monroy2025gr2r}.

Deep restoration models have substantially advanced reconstruction quality by learning expressive priors from the data~\cite{dpir,diffpir}. Plug-and-play (PnP~\cite{venkatakrishnan2013plug}) and related regularization-by-denoising (RED~\cite{romano2017red}) frameworks further decouple priors from forward models by replacing handcrafted regularizers with learned denoisers, enabling reuse across operators and applications~\cite{venkatakrishnan2013plug, romano2017red}. Unrolled networks inspired by convex optimization methods such as the Alternating Direction Method of Multipliers (ADMM~\cite{10.1561/2200000016}) and Half Quadratic Splitting (HQS~\cite{geman1995nonlinear})  go a step further by embedding physics-driven data-consistency steps into trainable iterative solvers, improving interpretability and operator-awareness~\cite{yonina,aggarwal2018modl}. Nevertheless, robustness remains a persistent limitation. Many learned priors operate directly in pixel space and are optimized for a narrow training distribution; operator-induced distortions and distribution shifts can therefore cause brittle behavior, especially when supervision is removed or when noise is non-Gaussian and signal-dependent~\cite{tachella2022unsupervised,monroy2025gr2r}. These challenges motivate priors to act in more stable representation spaces, where the semantic structure is preserved between measurement operators, noise regimes, and datasets~\cite{cheng2024transfer,lin2022frozen,zhang2024frozen}.

In parallel, the computer vision community has observed rapid progress in foundation encoders trained at web scale, producing transferable representations that support broad downstream generalization~\cite{clip,lin2022frozen,zhang2024frozen}. Vision--language pretraining with CLIP is a prominent example, learning content-related and distortion-invariant features from large-scale image--text data~\cite{clip}. Importantly, recent evidence indicates that such representations are not only useful for recognition, but can also serve as effective building blocks for restoration and other dense prediction tasks: Transfer CLIP~\cite{cheng2024transfer} shows that frozen CLIP with Resnet-50 (RN50) dense multi-scale features remain remarkably consistent under common corruptions and can drive strong out-of-distribution denoising with a lightweight decoder, and related works report successful transfer of frozen pretrained representations for video understanding~\cite{lin2022frozen} and semantic segmentation~\cite{zhang2024frozen}. This suggests a promising direction for inverse problems: instead of learning priors from scratch (or heavily fine-tuning large models), foundation encoders can provide a robust, task-agnostic representational backbone, while compact task adapters learn the minimal transformation needed for reconstruction~\cite{cheng2024transfer,ram}.

\begin{figure}[t]
    \centering
    \includegraphics[width=\linewidth]{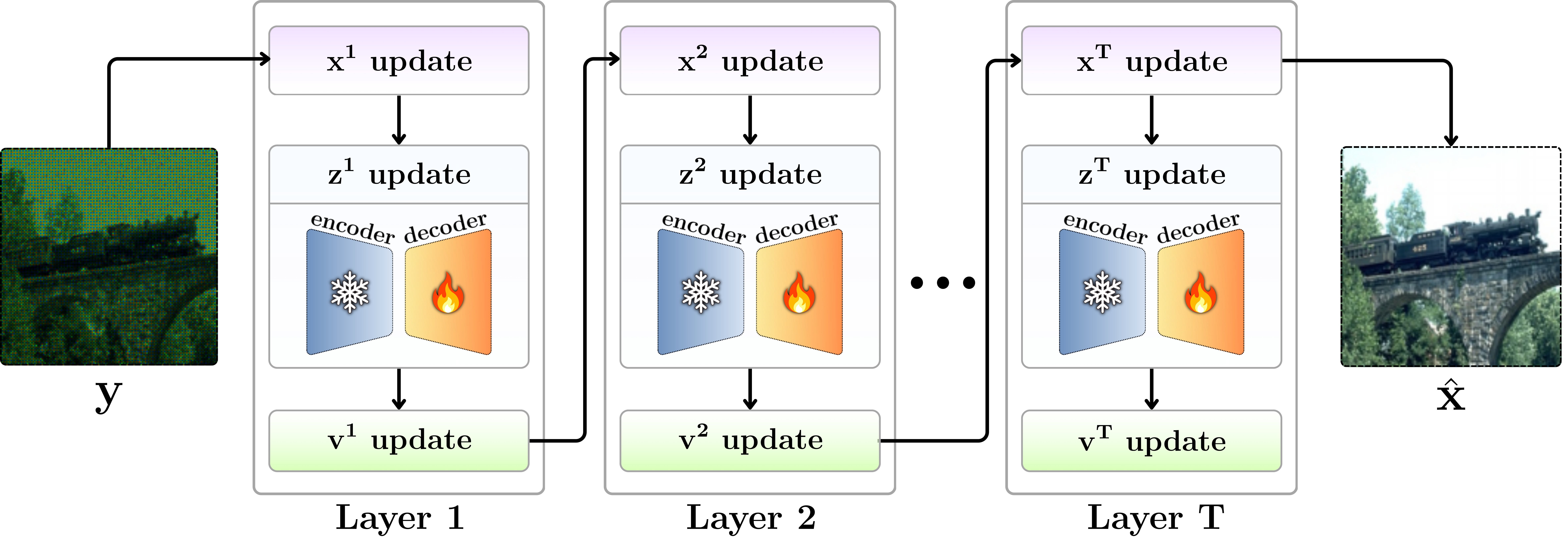} 
    \caption{\textbf{Overview of the proposed method.} The reconstruction network with $T$ iterations alternates between: (i) a data-consistency update $\mathbf{x}^t$, and (ii) a CLIP-based prior update $\mathbf{z}^t$. The prior consists of a frozen CLIP encoder coupled with a learnable decoder. After $T$ iterations, the updates yield the reconstruction $\hat{\mathbf{x}}$.}
    \label{fig:main}
\end{figure}

This paper proposes a way to turn a strong computer-vision foundation model into a practical prior for Poisson inverse problems. The key idea is to anchor restoration on frozen CLIP RN50 dense multi-scale features, treating them as a stable representation space that transfers across operators and noise regimes, and to couple this prior with an ADMM-inspired unrolled solver for principled, operator-aware data consistency. Beyond supervised training, a robust self-supervised learning scheme is formulated for signal-dependent Poisson noise by integrating Generalized Recorrupted-to-Recorrupted (GR2R) objectives in the measurement domain~\cite{monroy2025gr2r} with an Equivariant Imaging regularizer that stabilizes learning under sampling via virtual acquisitions~\cite{chen2021equivariant}. The resulting framework (Fig.~\ref{fig:main}) targets both reconstruction quality and robustness under dataset and acquisition shifts, with particular emphasis on Poisson CFA demosaicing and Poisson deblurring. The main contributions are summarized as follows:

\begin{itemize}
    \item A foundation-model prior for Poisson inverse problems is introduced by integrating a \emph{frozen} CLIP image encoder with a lightweight trainable decoder within an ADMM-inspired unrolled PnP solver, enabling parameter-efficient adaptation while preserving representation stability.
    \item A self-supervised training scheme specialized to signal-dependent Poisson noise is proposed by coupling GR2R measurement-domain re-corruption with Equivariant Imaging regularization to stabilize learning under sampling operators and prevent degenerate solutions.
    \item Empirical results on Poisson CFA demosaicing and Poisson deblurring demonstrate competitive reconstruction quality, improved robustness under dataset shifts, and favorable efficiency in terms of reconstruction time and TFLOPs, compared to iterative PnP baselines.
\end{itemize}

\section{Image Formation}
\label{sec:image_formation}

Let $\mathbf{x}\in\mathbb{R}^{3n}$ denote the unknown clean RGB image (vectorized), with spatial dimensions $n = H\times W$ and three color channels. Let $\mathbf{y}\in\mathbb{R}^{m}$ denote the observed measurements. Is considered a photon-limited acquisition model with Poisson statistics,
\begin{equation}
\mathbf{y}
=
\gamma \cdot \mathrm{Poisson}\!\left(\mathbf{A}\mathbf{x}/\gamma\right),
\label{eq:meas_model}
\end{equation}
where $\mathbf{A}\in\mathbb{R}^{m\times 3n}$ is a known linear forward operator, $\mathrm{Poisson}(\cdot)$ \mila{denotes a Poisson process and acts element-wise}, and $\gamma>0$ controls the noise severity (the higher $\gamma$ yields a higher shot-noise level). To facilitate a precise description of each inverse problem, $\mathbf{A}$ is specialized as follows:

\paragraph{Poisson demosaicing (CFA sampling).}
In demosaicing, $\mathbf{A}$ models color filter array (CFA) subsampling. Let \mila{$\mathbf{x}=[\mathbf{x}_R;\mathbf{x}_G;\mathbf{x}_B]$} with $\mathbf{x}_c\in\mathbb{R}^{n}$ for $c\in\{R,G,B\}$. The Bayer BGGR acquisition can be written as:
\begin{equation}
\mathbf{A}\mathbf{x}
=
\mathbf{M}_R \mathbf{x}_R
+
\mathbf{M}_G \mathbf{x}_G
+
\mathbf{M}_B \mathbf{x}_B,
\label{eq:cfa_forward}
\end{equation}
where $\mathbf{M}_R,\mathbf{M}_G,\mathbf{M}_B\in \{0,1\}^{n\times n}$ are diagonal binary masking matrices encoding the CFA pattern (with disjoint supports and $\mathbf{M}_R+\mathbf{M}_G+\mathbf{M}_B=\mathbf{I}$). Consequently, $m=HW$ and $\mathbf{y}\in\mathbb{R}^{n}$ is a single-channel mosaiced observation corrupted by Poisson noise.

\paragraph{Poisson deblurring (spatial convolution).}
In deblurring, $\mathbf{A}$ applies the same spatial blur to each color channel. Let $\mathbf{H}\in\mathbb{R}^{n \times n}$ denote the linear convolution operator induced by a blur kernel $\mathbf{h} \in \mathbb{R}^{k \times k}$ (e.g., under periodic boundary conditions), where $k$ is the kernel size, so that $\mathbf{H}\mathbf{x}_c=\mathbf{h}*\mathbf{x}_c$, where $*$ denotes spatial convolution. Therefore, the forward model is formulated as:
\begin{equation}
\mathbf{A}\mathbf{x}
=
\begin{bmatrix}
\mathbf{H}\mathbf{x}_R\\
\mathbf{H}\mathbf{x}_G\\
\mathbf{H}\mathbf{x}_B
\end{bmatrix},
\label{eq:deblur_forward}
\end{equation}
denoting a channel-wise convolution with a shared kernel. Hence $m=3HW$ and $\mathbf{y}$ is a blurred noisy RGB image following the Poisson model.

\section{Method}
\label{sec:method}

Given the observed measurements $\mathbf{y}$ and the forward operator $\mathbf{A}$, the restoration task is formulated as a regularized inverse problem:
\begin{equation}
\widehat{\mathbf{x}}
= \underset{\mathbf{x}}{\arg \min}
\ \frac{1}{2}\big\| \mathbf{A}\mathbf{x}-\mathbf{y}\big\|_2^2
+\lambda\,\mathcal{R}(\mathbf{x}),
\label{eq:reg_obj}
\end{equation}
where the first term enforces data-fidelity and, for computational efficiency, is implemented as a quadratic $\ell_2$ penalty\footnotemark, and $\mathcal{R}(\cdot)$ encodes prior information weighted by $\lambda>0$.

\footnotetext{A quadratic data-fidelity surrogate to match the training objectives and enable closed-form data-consistency updates. For supervised, the loss measures estimation error against clean targets. For self-supervised, the Poisson statistics are handled implicitly by GR2R re-corruption applied in the measurement domain of each task.}

\newpage
\subsection{Unrolling Optimization.}
To obtain an interpretable restoration architecture, Equation~\eqref{eq:reg_obj} is solved using an ADMM scheme, whose iterations can be unrolled into a finite-depth feed-forward network. Specifically, this scheme separates the problem into two complementary updates: (i) a \emph{data-consistency} step associated with the fidelity term, and (ii) a \emph{regularization} step associated with the prior. In practice, the Poisson likelihood is handled through a quadratic data-fidelity surrogate in Equation~\eqref{eq:reg_obj}, which enables a simple ADMM splitting with a closed-form data-consistency step. Introducing an auxiliary variable $\mathbf{z}$ and a scaled dual variable $\mathbf{u}$, the scaled-form augmented Lagrangian becomes:
\begin{equation}
\mathcal{L}_\rho(\mathbf{x},\mathbf{z},\mathbf{u})
=
\frac{1}{2}\big\| \mathbf{A}\mathbf{x}-\mathbf{y}\big\|_2^2
+\lambda\,\mathcal{R}(\mathbf{z})
+\frac{\rho}{2}\big\|\mathbf{x}-\mathbf{z}+\mathbf{u}\big\|_2^2
-\frac{\rho}{2}\|\mathbf{u}\|_2^2,
\label{eq:scaled_augLag}
\end{equation}
with penalty parameter $\rho>0$. Minimizing Equation~\eqref{eq:scaled_augLag} alternately with respect to $\mathbf{x}$ and $\mathbf{z}$, followed by a dual ascent step, yields the following iterations:
\begin{subequations}
\label{eq:admm_updates}
\begin{align}
\mathbf{x}^{t+1}
&=
\underset{\mathbf{x}}{\arg \min}
\ \frac{1}{2}\big\|\mathbf{A}\mathbf{x}-\mathbf{y}\big\|_2^2
+
\frac{\rho}{2}\big\|\mathbf{x}-\mathbf{z}^{t}+\mathbf{u}^{t}\big\|_2^2,
\label{eq:x_update}\\
\mathbf{z}^{t+1}
&=
\underset{\mathbf{z}}{\arg \min}
\ \lambda\,\mathcal{R}(\mathbf{z})
+
\frac{\rho}{2}\big\|\mathbf{x}^{t+1}-\mathbf{z}+\mathbf{u}^{t}\big\|_2^2,
\label{eq:z_update_argmin}\\
&= \text{prox}_{\frac{\lambda}{\rho}\mathcal{R}}\!\big(\mathbf{x}^{t+1}+\mathbf{u}^t\big),
\notag\\
\mathbf{u}^{t+1}
&=
\mathbf{u}^{t}+\mathbf{x}^{t+1}-\mathbf{z}^{t+1},
\label{eq:u_update}
\end{align}
\end{subequations}
where $\mathbf{u}$ denotes the scaled dual variable. This subproblem splitting offers a natural framework for incorporating a PnP prior as solution for Equation~\eqref{eq:z_update_argmin}, instantiated by a CLIP-based denoising prior, which is described in detail in the subsequent subsection.

\subsection{Data Consistency.}
The $\mathbf{x}$-subproblem in Equation~\eqref{eq:x_update} enforces measurement fidelity by updating the current estimate so that it matches the observation $\mathbf{y}$ under the forward operator $\mathbf{A}$, while remaining close to the auxiliary variable through the quadratic penalty. This step solves a strictly convex quadratic problem, whose optimality condition yields the normal equation:
\begin{equation}
\big(\mathbf{A}^{\top}\mathbf{A}+\rho\,\mathbf{I}\big)\mathbf{x}^{t+1}
=
\mathbf{A}^{\top}\mathbf{y}
+\rho\big(\mathbf{z}^{t}-\mathbf{u}^{t}\big).
\label{eq:dc_normal_equation}
\end{equation}
Importantly, for the inverse problems considered in this work, Equation~\eqref{eq:dc_normal_equation} admits closed-form solutions due to the structure of $\mathbf{A}$.

\paragraph{Closed-form for demosaicing (CFA).}
In demosaicing, $\mathbf{A}$ is the CFA mosaicing operator, hence $\mathbf{A}^{\top}\mathbf{A}$ is diagonal with entries $\big(\mathbf{A}^{\top}\mathbf{A}\big)_{ii}\in\{0,1\}$. Therefore, Equation~\eqref{eq:dc_normal_equation} decouples element-wise, leading to the following update:
\begin{equation}
\big(\mathbf{x}^{t+1}\big)_i =
\frac{\big(\mathbf{A}^{\top}\mathbf{y}\big)_i+\rho\big(\mathbf{z}^{t}-\mathbf{u}^{t}\big)_i}
{\big(\mathbf{A}^{\top}\mathbf{A}\big)_{ii}+\rho},
\qquad i=1,\dots,3n.
\label{eq:x_update_elemwise}
\end{equation}
This update can be computed as the per-pixel division on the RGB grid.

\paragraph{Closed-form for deblurring (convolution).}
For fast image deblurring, we can take advantage of the block-circulant structure of the forward operator $\mathbf{A}$ that is diagonalizable in the Fourier domain. Under this standard assumption, let $\mathcal{F}(\cdot)$ and $\mathcal{F}^{-1}(\cdot)$ denote the (discrete) Fourier transform and its inverse. Then the normal equation in~\eqref{eq:dc_normal_equation} admits the following closed-form channel-wise update, efficiently computed via FFTs:
\begin{equation}
\mathbf{x}^{t+1}
=
\mathcal{F}^{-1}\!\left[
\frac{
\overline{\mathcal{F}(\mathbf{h})}\,\mathcal{F}(\mathbf{y})
+\rho\,\mathcal{F}\!\big(\mathbf{z}^{t}-\mathbf{u}^{t}\big)
}{
\lvert \mathcal{F}(\mathbf{h}) \rvert^{2}
+\rho
}
\right].
\label{eq:x_update_deblur_fft}
\end{equation}
Consequently, the data-consistency step remains computationally efficient, requiring only pointwise operations in the Fourier domain.

\subsection{CLIP Denoising Prior.} The $\mathbf{z}$-subproblem in Equation~\eqref{eq:z_update_argmin} corresponds to the regularization step of the ADMM algorithm. Updates $\mathbf{z}$ by balancing two effects: staying close to the current estimate $\mathbf{x}^{t+1}+\mathbf{u}^t$ and promoting the prior $\mathcal{R}(\mathbf{z})$. This update can be viewed as a denoising/proximal operation applied to $\mathbf{x}^{t+1}+\mathbf{u}^t$, with its strength controlled by $\lambda/\rho$. In a PnP setting, $\mathcal{R}$ does not need to be written explicitly; instead, the solution of Equation~\eqref{eq:z_update_argmin} is approximated by a denoiser $\mathcal{D}_\theta(\cdot)$. The denoiser is defined as the composition of an encoder $\mathcal{E}_{\text{CLIP}}$ and decoder $\mathcal{G}_\theta$ as follows:

\begin{equation}
\mathbf{z}^{t+1} 
    \\ = 
    \mathcal{D}_\theta( \tilde{\mathbf{z}})
= \mathcal{G}_{\theta}\!\Big(\mathcal{E}_{\text{CLIP}}( \tilde{\mathbf{z}})\Big),
\label{eq:clip_comp}
\end{equation}
where $\tilde{\mathbf{z}}=\mathbf{x}^{t+1}+\mathbf{u}^{t}$ is the input to the prior step, $\mathcal{E}_{\text{CLIP}}(\cdot)$ is the encoder, 
and $\mathcal{G}_{\theta}(\cdot)$ is the decoder that maps CLIP features back to an RGB estimate.
The encoder parameters are kept fixed during training, this choice is well motivated because, as presented by Cheng et al. in~\cite{cheng2024transfer}, frozen CLIP ResNet \emph{dense multi-scale} features exhibit two desirable properties used in restoration: \textbf{distortion-invariant}, features extracted from a clean image and from its distorted/noisy versions remain highly similar across corruption levels (i.e., the representation is stable to low-level degradations), and \textbf{content-related}, the same features still organize the embedding space primarily by underlying semantics/content, preserving identity/structure despite corruption. Distortion invariance prevents the prior step from fitting measurement-induced artifacts, while content-relatedness preserves the latent scene structure that should be reconstructed, enabling $\mathcal{G}_\theta$ to map robust CLIP features back to a plausible RGB estimate.

Keeping $\mathcal{E}_{\text{CLIP}}$  frozen is crucial to retain these pretrained invariances (which can degrade under task-specific fine-tuning). Moreover, Cheng et al. report that the transfer of CLIP features provides strong denoising performance and improved generalization~\cite{cheng2024transfer}. Based on these findings, the same inductive bias can be effectively leveraged in broader inverse problems: using $\mathcal{D}_\theta$ as solver of the $\mathbf{z}$-subproblem, the prior step repeatedly projects iterate onto a robust, content-preserving manifold, improving generalization beyond denoising.

Following Transfer CLIP~\cite{cheng2024transfer}, $\mathcal{E}_{\text{CLIP}}$ is instantiated with the CLIP RN50 visual backbone. Dense multi-scale feature maps are extracted from intermediate stages (before global pooling), preserving spatial resolutions that progressively downsample the input (approximately from $H/2$ down to $H/16$). These features are fed to a lightweight U-Net-like decoder $\mathcal{G}_{\theta}$, which upsamples the coarsest map and fuses information across scales via skip connections. Each decoding stage uses standard convolutional blocks (Conv+ReLU), and a final $3\times3$ convolution produces the RGB estimate.

\subsection{Unrolling Network.}
The proposed reconstructor is implemented as a finite-depth unrolled architecture obtained by truncating \(T\) iterations of the ADMM updates presented in Equation~\eqref{eq:admm_updates} and denoted by \(\mathcal{F}_t\). At each \(t\)-th iteration, the ADMM variables are
\(
\chi^t=\{\mathbf{x}^{t},\mathbf{z}^{t},\mathbf{u}^{t}\}.
\)
Then, the forward pass can be written as the composition of multiple blocks and can be compactly written as:
\begin{equation}
\hat{\mathbf{x}}
= f_\theta(\mathbf{y}; \mathbf{A}) = 
\mathcal{F}_T \circ \cdots \circ \mathcal{F}_1\!(\mathbf{y}, \chi^0;\theta, \mathbf{A}),
\label{eq:unroll_comp}
\end{equation}  where \(f_\theta\) represents the unrolled network with trainable parameters $\theta$ which maps noisy measurements $\textbf{y}$ to the RGB estimation. It is worth highlighting that these parameters correspond to the decoder weights $\mathcal{G}_\theta$, which are shared across all iterations, the ADMM hyperparameter $\rho$ and $\lambda$ are fixed. For initialization, the state is set to \(\chi^0 = \{ \mathbf{A}^\top\mathbf{y}, \mathbf{0}, \mathbf{0}\}\).
The choice \(\mathbf{x}^0=\mathbf{A}^\top\mathbf{y}\) corresponds to a back-projection of the mosaiced measurements to the RGB grid, providing a simple physics-driven warm-start. The auxiliary and dual variables are initialized to zero, which is standard in scaled ADMM and avoids introducing additional bias at the first stage.

\subsection{Training Objectives.}

\subsubsection{Supervised training.}  In  this setting, paired clean targets are available, so the network can be optimized directly for reconstruction accuracy. Given paired data $(\mathbf{y},\mathbf{x})$, the supervised objective is:
\begin{equation}
\mathcal{L}_{\mathrm{sup}}(\theta)
=
\left\|f_{\theta}(\mathbf{y};\mathbf{A})-\mathbf{x}\right\|_2^2.
\label{eq:lsup}
\end{equation} 
This loss directly penalizes the reconstruction error against the clean target, allowing the network to learn both an inversion of the mosaicing process and an implicit image prior from data. In the \textit{self-supervised} setting, clean ground truth is not available (e.g., photon-limited imaging), and training must rely only on noisy mosaiced measurements and the known forward model in Equation~\eqref{eq:meas_model}.

\subsubsection{Self-supervised training.} In this setting, two complementary objectives for minimization are proposed: the first aims to reduce the measurement consistency error associated with noisy observations through the GR2R self-supervised loss $\mathcal{L}_{\text{GR2R}}$~\cite{monroy2025gr2r}, while the second aims to prevent degenerate solutions by enforcing learning beyond the range space of the forward operator $\mathbf{A}$ through an Equivariant Imaging (EI) regularization term $\mathcal{L}_{\text{EI}}$. 

GR2R is adopted for self-supervision since Poisson noise is \emph{signal-dependent} and non-Gaussian, so standard self-supervised losses designed for additive Gaussian noise are not appropriate. In addition, an equivariant-imaging term is included to stabilize training under CFA sampling, which may otherwise admit trivial solutions when only measurement consistency is enforced. Consequently, the overall self-supervised loss function is defined as follows:

\begin{equation}
    \mathcal{L}_{\text{self}}(\mathbf{y}; \theta) = \mathcal{L}_{\text{GR2R}}(\mathbf{y};\theta) + \tau \mathcal{L}_{\text{EI}}(\mathbf{y};\theta),
\end{equation} 
where $\tau>0$ is a 
weighty parameter. Here, $f_\theta(\cdot;\mathbf{A})$ maps inputs $\mathbf{y}$ to RGB estimation $\hat{\mathbf{x}}$, while consistency is enforced in the measurement domain through $\mathbf{A}$.
 The GR2R~\cite{monroy2025gr2r} loss, considering the forward operator $\mathbf{A}$, is defined as: \begin{equation}
    \mathcal{L}_\text{GR2R}(\mathbf{y},\theta)=\Vert \mathbf{A}f_\theta(\mathbf{y}_1;\mathbf{A})-\mathbf{y}_2\Vert_2^2,
\end{equation}
here, the recorrupted pair $(\mathbf{y}_1,\mathbf{y}_2)$ is constructed in the measurement domain, i.e., $\mathbf{y}_1,\mathbf{y}_2\in\mathbb{R}^{n}$, so the loss follows the Poisson statistics of the measurements. For Poisson noise, these pairs can be constructed as follows: \begin{equation}
    \begin{aligned}
        \mathbf{y}_1 = \frac{\mathbf{y} - \gamma \mathbf{w}}{1 - \alpha}, \quad 
        \mathbf{y}_2 = \frac{1}{\alpha}\mathbf{y} - \frac{1-\alpha}{\alpha} \mathbf{y}_1,
    \end{aligned} 
\end{equation} 
where $\mathbf{w} \sim \text{Bin}( \mathbf{y}/\gamma, \alpha)$, and $\alpha  \in (0,1)$ controls the level of re-corruption on each pair and $\text{Bin}(\cdot)$ denotes the binominal distribution. On the other hand, the equivariant imaging regularization term $\mathcal{L}_{\text{EI}}$ is introduced to enable learning beyond the range space of the forward operator $\mathbf{A}$~\cite{chen2021equivariant}. Specifically, this regularization induces a virtual data-augmentation mechanism in network’s estimations to simulate new acquisitions and to enforce an equivariance constraint with respect to a prescribed set of transformations. This regularizer is defined as follows:
\begin{equation}
\begin{aligned}
    &\mathcal{L}_{\text{EI}}(\mathbf{y};\theta) = \Vert \mathbf{x}_{v} - f_\theta( \mathbf{y}_v;\mathbf{A})\Vert_2^2, \\
    & \mathbf{y}_v = \mathbf{A}\mathbf{x}_v, \quad \mathbf{x}_v = \mathcal{T}(\hat{\mathbf{x}}).
\end{aligned}
\end{equation}
Observe the “virtual’’ acquisition $\mathbf{y}_v$ is obtained by simulating the response to a transformed version $\mathbf{x}_v$ of the current estimate $\hat{\mathbf{x}} = f_\theta(\mathbf{y}; \mathbf{A})$, under a prescribed transformation group $\mathcal{T}$.
In this work, the effective set of transformations used to generate these virtual samples is chosen according to the forward operator: for Poisson demosaicing, random rotations are adopted, motivated by the spatially periodic structure of CFAs; for Poisson deblurring, random scaling transformations are employed following the procedure proposed by authors in~\cite{scanvic2026scale}.

\section{Simulations and Results}
\label{sec:sim_results}

The proposed self-supervised approach is compared against representative PnP and modern restoration baselines: DPIR~\cite{dpir}, GSPnP~\cite{gspnp}, and RAM~\cite{ram}. In addition, Transfer CLIP~\cite{venkatakrishnan2013plug, cheng2024transfer} is included as a semantic-prior baseline. 
DPIR~\cite{dpir} performs PnP restoration by embedding a pretrained deep CNN denoiser as an implicit prior within an iterative optimization scheme. 
GSPnP~\cite{gspnp} introduces a convergent PnP formulation that alternates a data consistency gradient step with a denoising operator to improve stability. 
RAM~\cite{ram} is a lightweight foundation model for computational imaging that supports broad reconstruction tasks using a single pretrained network. 
Transfer CLIP~\cite{venkatakrishnan2013plug, cheng2024transfer} incorporates a CLIP-based semantic prior into a PnP procedure; the non-finetuned setting is evaluated, \ie, the CLIP prior is used as-is without any task-specific fine-tuning.

\subsection{Experimental Setup}

\noindent
\textbf{Datasets.} Image demosaicing under Poisson noise is evaluated on two widely used natural-image benchmarks: BSDS500 and DIV2K. 
Following standard practice, BSDS500 is split into 200 training images, 100 validation images, and 200 test images. 
During training on BSDS500, images are randomly cropped into patches of size $256\times256$. To assess generalization, results are additionally reported on the DIV2K test set using models trained on BSDS500; for evaluation, DIV2K images are center-cropped to $256\times256$.

\noindent
\textbf{Noise model.} Poisson noise is used according to the measurement model in Equation~\eqref{eq:meas_model}, with noise severity controlled by the scaling parameter $\gamma\in\{0.01,0.05\}$, where larger $\gamma$ corresponds to a more challenging photon-limited regime.

\noindent
\textbf{Training setup.}
The CLIP denoiser module is initialized from the pretrained decoder weights of~\cite{cheng2024transfer}; only the decoder is updated during training, while the RN50 CLIP image encoder remains frozen. Two variants are evaluated using $\mathcal{L}_{\text{self}}$ and $\mathcal{L}_{\text{sup}}$. For self-supervised training, the GR2R parameter is fixed to $\alpha=0.2$ and the EI weight to $\tau=0.1$. Models are trained for 2000 epochs with batch size 32, learning rate $10^{-4}$, and Adam. The unrolling depth is $T=2$ for demosaicing and $T=3$ for deblurring, with parameters shared across iterations. Deblurring uses a $9\times9$ Gaussian kernel with standard deviation $(1,1)$ pixels and periodic boundary conditions. For the proposed ADMM updates and applicable baselines, $\lambda=1.0$ and $\rho$ is tuned by random grid search over $[0.01,1.0]$.

\begin{table}[!t]
\centering
\caption{Demosaicing performance under Poisson noise. \vspace{-1em}}
\label{tab:poisson_noise}
\small
\setlength{\tabcolsep}{4pt}
\renewcommand{\arraystretch}{1.10}
\resizebox{\linewidth}{!}{%
\begin{tabular}{l c cc|cc}
\hline
\multirow{3}{*}{\textbf{Method}} &  & \multicolumn{4}{c}{\textbf{Poisson Noise}} \\
\cline{3-6}
& \multirow{2}{*}{\textbf{$\gamma$}} &
\multicolumn{2}{c|}{\textbf{BSDS500}} & \multicolumn{2}{c}{\textbf{DIV2K}} \\
\cline{3-6}
& & \textbf{PSNR [dB]} & \textbf{SSIM} & \textbf{PSNR [dB]} & \textbf{SSIM} \\
\hline
DPIR~\cite{dpir}      & \multirow{6}{*}{\textbf{0.01}} & 25.96$\pm$1.98 & 0.7039$\pm$0.0887 & 25.95$\pm$2.82 & 0.7093$\pm$0.1244 \\
Transfer CLIP~\cite{cheng2024transfer}  &                                 & 27.88$\pm$3.18 & 0.7833$\pm$0.1068 & 27.50$\pm$4.44 & 0.7841$\pm$0.1006 \\
GSPnP~\cite{gspnp}    &                                 & 29.25$\pm$2.16 & 0.8061$\pm$0.0647 & 29.36$\pm$2.62 & 0.8237$\pm$0.0605 \\
RAM~\cite{ram}        &                                 & 30.46$\pm$2.07 & \textbf{0.8622$\pm$0.0545} & \textbf{30.48$\pm$2.51} & \textbf{0.8729$\pm$0.0568} \\
\textbf{Ours (Self)}  &                                 & \textbf{30.53$\pm$2.11} & 0.8609$\pm$0.0476 & 30.30$\pm$2.64 & 0.8641$\pm$0.0537 \\
\textbf{Ours (Sup)}  &                                 & \textbf{30.75$\pm$2.13} & \textbf{0.8703$\pm$0.0458} & \textbf{30.56$\pm$2.65} & \textbf{0.8735$\pm$0.0488} \\
\hline
DPIR~\cite{dpir}      & \multirow{6}{*}{\textbf{0.05}} & 24.71$\pm$2.08 & 0.6358$\pm$0.0951 & 24.55$\pm$2.81 & 0.6397$\pm$0.1242 \\
Transfer CLIP~\cite{cheng2024transfer}   &                                 & 21.91$\pm$3.14 & 0.5716$\pm$0.1154 & 21.67$\pm$5.22 & 0.6122$\pm$0.1179 \\
GSPnP~\cite{gspnp}    &                                 & 26.50$\pm$2.21 & 0.6945$\pm$0.1007 & 26.58$\pm$2.77 & 0.7193$\pm$0.1057 \\
RAM~\cite{ram}        &                                 & 26.17$\pm$1.65 & 0.7178$\pm$0.0699 & 26.33$\pm$2.24 & 0.7343$\pm$0.0721 \\
\textbf{Ours (Self)}  &                                 & \textbf{26.98$\pm$1.74} & \textbf{0.7579$\pm$0.0656} & \textbf{26.60$\pm$2.58} & \textbf{0.7435$\pm$0.0744} \\
\textbf{Ours (Sup)}  &                                 & \textbf{27.04$\pm$2.07} & \textbf{0.7522$\pm$0.0744} & \textbf{26.91$\pm$2.64} & \textbf{0.7669$\pm$0.0712} \\
\hline
\end{tabular}%
}
\end{table}

\begin{figure*}[!t]
    \centering
    \includegraphics[width=\textwidth]{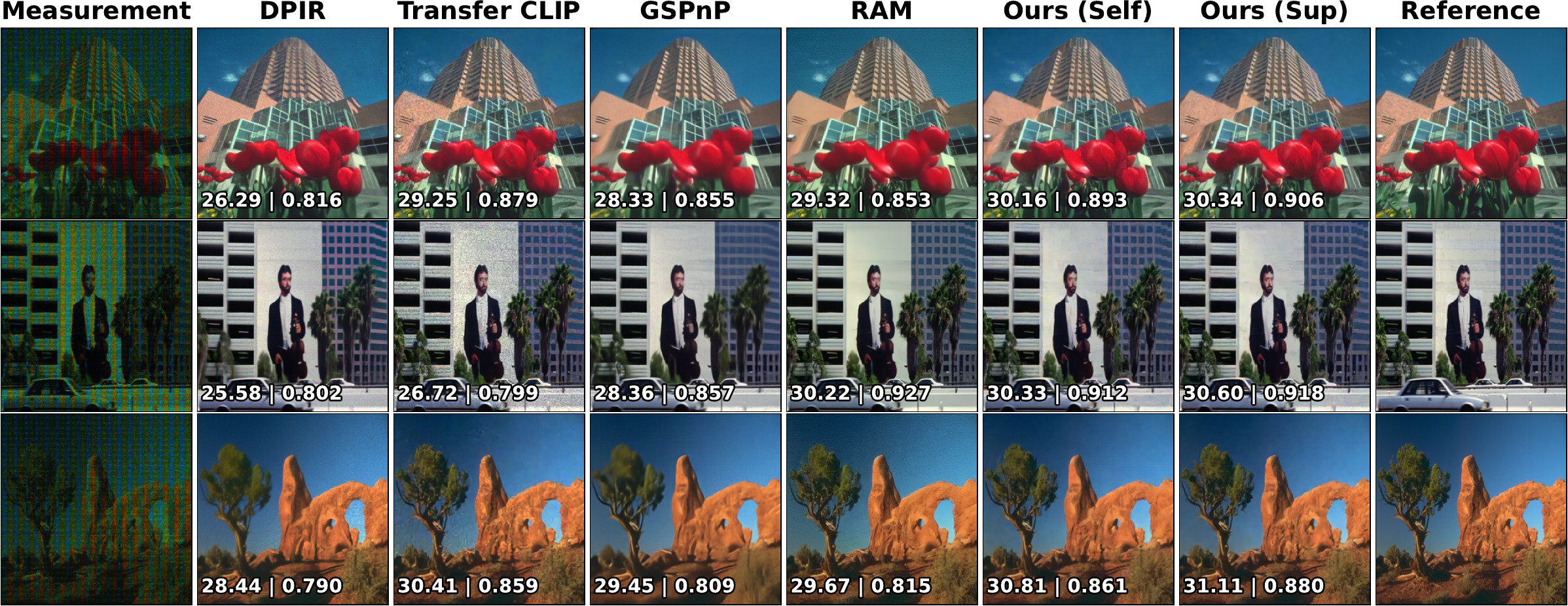} \vspace{-2em}
    \caption{
    Qualitative comparison on BSDS500 for Poisson demosaicing with shot-noise level $\gamma=0.01$.
    Columns show the mosaiced noisy \emph{Measurement}, reconstructions by DPIR~\cite{dpir}, Transfer CLIP~\cite{venkatakrishnan2013plug,cheng2024transfer}, GSPnP~\cite{gspnp}, and RAM~\cite{ram}, followed by the proposed method (Ours~(Self)) and its supervised counterpart (Ours~(Sup)), and the \emph{Reference}. PSNR/SSIM are overlaid for direct visual comparison.
    } 
    \label{fig:bsds500_demosaic_sigma001_qual}
\end{figure*}

\subsection{Poisson demosaicing.}

Table~\ref{tab:poisson_noise} shows that the proposed method remains competitive across photon regimes and under the BSDS500 to DIV2K dataset shift. Moreover, the self-supervised objective (GR2R+EI) recovers most of the supervised performance, indicating that it provides a reliable learning signal under signal-dependent Poisson noise and CFA sampling.

At mild noise ($\gamma=0.01$), Ours (Self) achieves 30.53\,dB on BSDS500, improving over the strongest baseline, RAM, by +0.07\,dB and remaining close to Ours (Sup), which obtains 30.75\,dB. Under the corresponding dataset shift, Ours (Self) obtains 30.30\,dB on DIV2K. Although this is slightly below RAM by 0.18\,dB, it remains competitive with the strongest baseline and improves over GSPnP by +0.94\,dB. Ours (Sup) achieves the best PSNR and SSIM in this mild-noise regime on both datasets.

At severe noise ($\gamma=0.05$), Ours (Self) attains 26.98\,dB on BSDS500, exceeding the strongest PSNR baseline, GSPnP, by +0.48\,dB. On DIV2K, Ours (Self) obtains 26.60\,dB, essentially tying GSPnP in PSNR (+0.02\,dB), while Ours (Sup) achieves the highest PSNR with 26.91\,dB. In this regime, the advantage is more pronounced in SSIM: on BSDS500, Ours (Self) improves over the best baseline SSIM from 0.7178 (RAM) to 0.7579, and on DIV2K from 0.7343 (RAM) to 0.7435.

As a closely related CLIP-based baseline, Transfer CLIP~\cite{cheng2024transfer} also relies on frozen CLIP features with a lightweight decoder, but operates without an explicit operator-aware data-consistency mechanism and is not trained with a Poisson-specific self-supervision signal; in Table~\ref{tab:poisson_noise}, the proposed unrolled solver consistently improves upon Transfer CLIP across both photon regimes and datasets.

Fig.~\ref{fig:bsds500_demosaic_sigma001_qual} reports a qualitative comparison on BSDS500 for Poisson demosaicing at $\gamma=0.01$.
The mosaiced measurements exhibit pronounced CFA-induced grid artifacts and color misplacement, which are further amplified by signal-dependent noise.
DPIR and Transfer CLIP partially suppress these degradations, but may retain residual zippering and chromatic inconsistencies, especially around thin structures and high-contrast edges. GSPnP typically produces cleaner restorations, yet mild high-frequency artifacts can persist. RAM yields visually pleasing outputs in several cases, although fine textures may be slightly attenuated, and subtle color shifts can appear.

\subsection{Poisson deblurring}

Table~\ref{tab:deblurring} reports results under Poisson noise in BSDS500 and DIV2K for two noise regimes ($\gamma\in\{0.01,0.05\}$).
Across all settings (two datasets $\times$ two $\gamma$ values), the proposed method (Ours) consistently attains the best performance, with the supervised variant delivering the top PSNR/SSIM and the self-supervised variant closely following.
This demonstrates that the approach is not tied to a specific forward operator (CFA mosaicing vs.\ blur) and transfers reliably across dataset distributions.

For Poisson deblurring, classical PnP baselines are already highly competitive at mild noise ($\gamma=0.01$), particularly GSPnP. In this setting, the PSNR margins are modest but consistent: relative to GSPnP, Ours (Self) improves by +0.14\,dB on BSDS500 and +0.07\,dB on DIV2K, while Ours (Sup) increases the margins to +0.43\,dB and +0.37\,dB, respectively. The SSIM gains are also consistent: Ours (Sup) improves over GSPnP by +0.023 on BSDS500 and +0.022 on DIV2K, suggesting improved structural fidelity beyond pixel-wise reconstruction accuracy.

At severe noise ($\gamma=0.05$), the advantage becomes clearer in both PSNR and SSIM. Compared to GSPnP, Ours (Self) improves PSNR by +0.42\,dB on BSDS500 and +0.49\,dB on DIV2K, while Ours (Sup) further increases the margins to +0.55\,dB and +0.67\,dB. Consistent SSIM improvements are also observed: Ours (Self) improves by +0.031/+0.034 on BSDS500/DIV2K, and Ours (Sup) by +0.039/+0.044.

\begin{table}[!t]
\centering
\caption{Deblurring performance under Poisson noise. \vspace{-1em}}
\label{tab:deblurring}
\small
\setlength{\tabcolsep}{4pt}
\renewcommand{\arraystretch}{1.10}
\resizebox{\linewidth}{!}{%
\begin{tabular}{l c cc|cc}
\hline
\multirow{3}{*}{\textbf{Method}} &  & \multicolumn{4}{c}{\textbf{Poisson Noise}} \\
\cline{3-6}
& \multirow{2}{*}{\textbf{$\gamma$}} &
\multicolumn{2}{c|}{\textbf{BSDS500}} & \multicolumn{2}{c}{\textbf{DIV2K}} \\
\cline{3-6}
& & \textbf{PSNR [dB]} & \textbf{SSIM} & \textbf{PSNR [dB]} & \textbf{SSIM} \\
\hline
DPIR~\cite{dpir}      & \multirow{6}{*}{\textbf{0.01}} & 28.81$\pm$2.60 & 0.7949$\pm$0.0699 & 28.25$\pm$3.66 & 0.7894$\pm$0.0816 \\
Transfer CLIP~\cite{cheng2024transfer}  &                                 & 24.63$\pm$2.73 & 0.7389$\pm$0.0605 & 24.33$\pm$3.75 & 0.7298$\pm$0.0932 \\
GSPnP~\cite{gspnp}    &                                 & 29.37$\pm$2.73 & 0.8149$\pm$0.0701 & 28.90$\pm$3.65 & 0.8143$\pm$0.0785 \\
RAM~\cite{ram}        &                                 & 28.67$\pm$2.57 & 0.8028$\pm$0.0667 & 28.40$\pm$3.35& 0.8078$\pm$0.0672 \\
\textbf{Ours (Self)}  &                                 & \textbf{29.51$\pm$2.56} & \textbf{0.8257$\pm$0.0647} & \textbf{28.97$\pm$3.59} & \textbf{0.8238$\pm$0.0678} \\
\textbf{Ours (Sup)}   &                                 & \textbf{29.80$\pm$2.72} & \textbf{0.8380$\pm$0.0650} & \textbf{29.27$\pm$3.80} & \textbf{0.8362$\pm$0.0695} \\
\hline
DPIR~\cite{dpir}      & \multirow{6}{*}{\textbf{0.05}} & 26.72$\pm$2.42 & 0.7113$\pm$0.0901 & 26.18$\pm$3.39 & 0.7085$\pm$0.0988 \\
Transfer CLIP~\cite{cheng2024transfer}   &                                 & 19.87$\pm$2.05 & 0.5339$\pm$0.0891 & 19.89$\pm$3.09 & 0.5478$\pm$0.1116 \\
GSPnP~\cite{gspnp}    &                                 & 27.03$\pm$2.46 & 0.7227$\pm$0.0908 & 26.51$\pm$3.31 & 0.7223$\pm$0.0971 \\
RAM~\cite{ram}        &                                 & 25.80$\pm$2.25 & 0.6792$\pm$0.0763 & 25.04$\pm$2.95 & 0.6679$\pm$0.0861 \\
\textbf{Ours (Self)}  &                                 & \textbf{27.45$\pm$2.56} & \textbf{0.7532$\pm$0.0888} & \textbf{27.00$\pm$3.68} & \textbf{0.7564$\pm$0.0932} \\
\textbf{Ours (Sup)}   &                                 & \textbf{27.58$\pm$2.52} & \textbf{0.7613$\pm$0.0870} & \textbf{27.18$\pm$3.59} & \textbf{0.7666$\pm$0.0897} \\
\hline
\end{tabular}%
}
\end{table}

\begin{figure*}[!t]
    \centering
    \includegraphics[width=\textwidth]{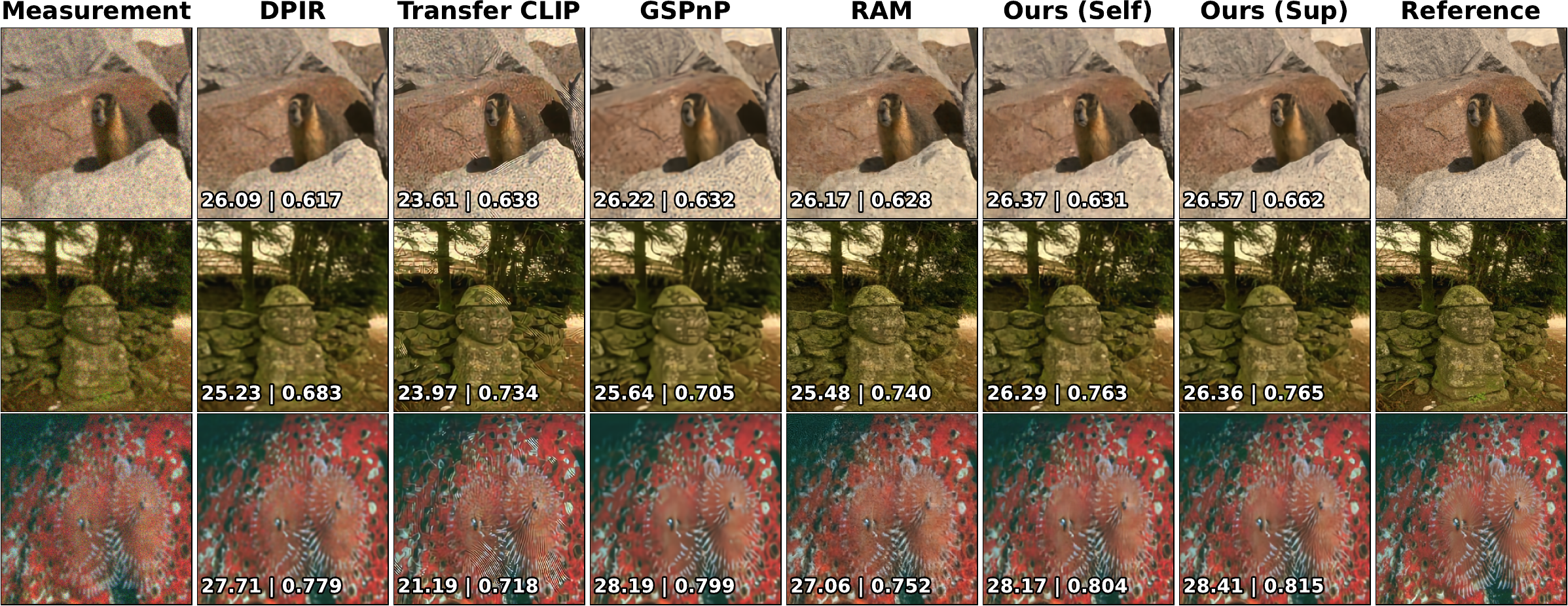} \vspace{-1.5em}
    \caption{
    Qualitative comparison on BSDS500 for Poisson deblurring with shot-noise level $\gamma=0.01$.
    Columns show the blurred noisy \emph{Measurement}, reconstructions by DPIR~\cite{dpir}, Transfer CLIP~\cite{venkatakrishnan2013plug,cheng2024transfer}, GSPnP~\cite{gspnp}, and RAM~\cite{ram}, followed by Ours~(Self), Ours~(Sup), and the \emph{Reference}.
    Per-image PSNR/SSIM are overlaid for direct visual comparison.
    }
    \label{fig:bsds500_deblur_sigma001_qual}
\end{figure*}

Fig.~\ref{fig:bsds500_deblur_sigma001_qual} shows the analogous comparison for Poisson deblurring at $\gamma=0.01$.
The blurred measurements lose high-frequency detail and exhibit noise-dependent grain, making texture recovery and edge localization challenging.
While DPIR, Transfer CLIP, and GSPnP recover plausible structures, they may leave residual blur or introduce ringing-like patterns in textured regions.
In the shown examples, RAM can be unstable under this setting and may introduce strong artifacts, which is reflected by the substantially degraded PSNR/SSIM overlays.
In contrast, the proposed method produces sharper contours with fewer structured artifacts, and the self-supervised variant remains consistently close to its supervised counterpart.

\break 

\paragraph{Additional observations.}
In both inverse problems, the gap between Ours (Sup) and Ours (Self) is systematic but moderate, indicating that the self-supervised objective captures a large fraction of the achievable performance while requiring no clean images. Finally, the non-finetuned baseline Transfer CLIP~\cite{cheng2024transfer} degrades substantially under stronger Poisson noise (and blur), highlighting that generalized pre-trained denoisers alone is insufficient when the data-fidelity is dominated by signal-dependent noise and the forward operator is ill-conditioned without task adaptation.

\subsection{Run Time and Complexity.}

Table~\ref{tab:benchmark} compares the evaluated methods in terms of model capacity, computational cost (TFLOPs, computed as denoiser complexity multiplied by the maximum number of iterations), and wall-clock inference time, all measured under a standardized setting (running on NVIDIA RTX 4070, input resolution $3\times256\times256$, 20 denoising realizations). As expected, iterative PnP pipelines coupled with heavy denoisers (e.g., DPIR, GSPnP, and Transfer CLIP) exhibit high compute and latency since the denoiser is repeatedly invoked across iterations (e.g., 9.00--11.49~TFLOPs and 0.52--2.47~s). While RAM substantially reduces compute (0.32~TFLOPs), it still incurs non-trivial runtime (0.8026~s), reflecting additional overhead beyond pure FLOP counts. In contrast, \textbf{Ours} attains a markedly lighter profile with \textbf{0.09}~TFLOPs and \textbf{0.0242}~s, yielding $\sim$100$\times$ speedup over Transfer CLIP and $\sim$20--40$\times$ over DPIR/GSPnP in this benchmark. Importantly, the reported \textbf{10.99M} parameters for \textbf{Ours} correspond to \emph{trainable parameters only} (i.e., the subset updated during training). The \emph{total} parameter count of the full pipeline remains comparable to Transfer CLIP, since both approaches rely on the same CLIP-based backbone; the key distinction is that the proposed method optimizes only a compact learnable component while keeping the remaining weights fixed/shared, which translates into substantially improved efficiency.

\begin{table}[!t]
\centering
\caption{Runtime and complexity comparison (test on size $3 \times 256 \times 256).$ \vspace{-1em}}
\label{tab:benchmark}
\small
\setlength{\tabcolsep}{4pt}
\renewcommand{\arraystretch}{1.10}
\resizebox{0.7\linewidth}{!}{%
\begin{tabular}{l r r r}
\hline
\textbf{Method} & \textbf{\#Params [M]} & \textbf{TFLOPs} & \textbf{Time [s]} \\
\hline
DPIR~\cite{dpir} & 32.64 & 11.49 & 0.5193 \\
Transfer CLIP~\cite{cheng2024transfer} & \textbf{10.99} & 9.00 & 2.4683 \\
GSPnP~\cite{gspnp} & 17.01 & 9.12 & 0.9247 \\
RAM~\cite{ram} & 34.13 & 0.32 & 0.8026 \\
\textbf{Ours} & \textbf{10.99} & \textbf{0.09} & \textbf{0.0242} \\
\hline
\end{tabular}%
}
\end{table}

\subsection{Ablation Study.}
\label{sec:ablation}

\begin{table}[h!]
\caption{Ablation study on training objectives and pretraining strategies. This ablation evaluates different backbones for the unrolled algorithm and different cost functions to assess their effectiveness on Poisson Demosaicing with $\gamma=0.01$. \vspace{-1em}} 
\label{tab:ablation}
\centering
\setlength{\tabcolsep}{5pt}
\renewcommand{\arraystretch}{1.10}
\resizebox{0.7\linewidth}{!}{%
\begin{tabular}{l l c c}
\hline
\textbf{Setup} & \textbf{Loss} & \textbf{PSNR [dB]} & \textbf{SSIM} \\
\hline
Untrained CLIP (Baseline) & $\mathcal{L}_{\text{sup}}$ & 21.91 & 0.7064\\
+ Pretraining &  $\mathcal{L}_{\text{MC}}$& 21.42 & 0.5134 \\
+ Pretraining & $\mathcal{L}_{\text{GR2R}}$ & 27.45 & 0.7992 \\
+ Pretraining &  EI~\cite{chen2021equivariant} & 28.41 & 0.7746 \\
+ Pretraining &  REI~\cite{chen2022robust} & 30.31 & 0.8600 \\
DnCNN & $\mathcal{L}_{\text{self}}$ & 29.90 & 0.8474 \\
+ Pretraining & $\mathcal{L}_{\text{self}}$ & 29.73 & 0.8459 \\
DRUnet & $\mathcal{L}_{\text{self}}$ & 29.82 & 0.8484 \\
+ Pretraining & $\mathcal{L}_{\text{self}}$ & 30.00 & 0.8543 \\
\textbf{Ours (Self)} & 
$\mathcal{L}_{\text{self}}$ & \textbf{30.53} & \textbf{0.8609} \\
\textbf{Ours (Sup)} & $\mathcal{L}_{\text{sup}}$  & \textbf{30.75} & \textbf{0.8703} \\
\hline
\end{tabular}  }
\end{table}

Table~\ref{tab:ablation} summarizes the impact of (i) the denoiser backbone used within the unrolled solver and (ii) the self-supervised training objective and initialization strategy. Two training regimes are considered: training from scratch for $3\,000$ epochs and fine-tuning from pretrained weights for $300$ epochs. Overall, training CLIP from scratch does not yield a meaningful prior: both supervised training and measurement-consistency-only optimization ($
\mathcal{L}_{\text{MC}} = \Vert \mathbf{A} f(\mathbf{y};\theta) - \mathbf{y} \Vert_2^2$) remain far from the best-performing configurations, indicating that data fidelity alone is insufficient and can lead to solutions that satisfy the forward model without recovering perceptually plausible structure. In contrast, switching to self-supervised objectives that promotes additional structure, such as GR2R and equivariance, consistently improves performance, with robust equivariant learning (REI~\cite{chen2022robust}) providing an alternative among the tested self-supervised cost functions. 

Beyond objectives, the backbone ablation shows that popular denoisers such as DnCNN and DRUNet are competitive under the same self-supervised setting, confirming that the gains are not merely an artifact of the unrolling architecture. Nevertheless, the key trend is that the use of a \emph{frozen} CLIP encoder enables substantially stronger reconstructions than attempting to learn CLIP features from scratch, supporting the hypothesis that pretrained CLIP representations provide robust content-related features that act as an effective prior for the inverse problem. Finally, combining the frozen CLIP encoder with the proposed self-supervised loss attains the best self-supervised performance in the table, while the supervised variant provides an upper bound that remains close, indicating that the proposed training strategy reduces the reliance on clean ground truth without sacrificing reconstruction quality\footnotemark.

\footnotetext{ Supplemental material, next page.  $\rightarrow \rightarrow \rightarrow$}

\section{Conclusion}
\label{sec:conclusion}
This work introduced a foundation-driven unrolled plug-and-play solver for photon-limited inverse imaging under Poisson noise. The proposed architecture unrolls a small number of ADMM iterations, combining (i) a closed-form data-consistency update tailored to the forward operator (CFA sampling or convolutional blur) with (ii) a parameter-efficient learned prior implemented as a lightweight decoder on top of \emph{frozen} CLIP RN50 dense multi-scale features. By freezing the encoder, the method preserves distortion-invariant and content-related representations, while restricting learning to a compact task adapter that is shared across unrolled iterations.
To enable learning without clean ground truth, a self-supervised objective was formulated by coupling GR2R measurement-domain re-corruption (to match signal-dependent Poisson statistics) with an Equivariant Imaging regularizer that stabilizes training under sampling and mitigates degenerate solutions. Experiments on Poisson CFA demosaicing and Poisson deblurring over BSDS500 and DIV2K demonstrated that the proposed approach achieves competitive quality and improved robustness under dataset shifts, with self-supervised performance approaching supervised training. In addition, the unrolled design and lightweight trainable prior yield favorable efficiency, substantially reducing computational cost and inference time compared to iterative PnP baselines that repeatedly employ heavy denoisers.

\section*{Acknowledgements}
The authors acknowledge the VIE of Universidad Industrial de Santander for supporting with “Apoyo a Semilleros de Investigación - Diseño de Codificación para el Muestreo Compresivo de Señales Multidimensionales Utilizando Técnicas Basadas en Aprendizaje Profundo”, Project code 4765.

\clearpage
\appendix

\begin{center}
{\sffamily\bfseries\huge\color{magenta!80!gray}
Supplemental Material\par}

\vspace{0.6em}

{\sffamily\bfseries\large
Frozen CLIP Priors for Robust\\
Self-Supervised Poisson Inverse Problems\par}

\vspace{1em}
{\color{magenta!80!gray}\hrule height 0.6pt}
\end{center}

\vspace{1.5em}

\section{Real Photon-Limited Data}
\label{sec:supp_real_sid}

To further assess the behavior of the proposed framework beyond synthetic Poisson simulations, an additional experiment is performed on real photon-limited data from the SID dataset~\cite{chen2018learning}. The experiment follows a zero-shot self-supervised adaptation protocol: a single real low-light image is used, one $512\times512$ patch is held out for testing, and the remaining patches are used only to optimize the self-supervised objective $\mathcal{L}_{\mathrm{self}}$. No clean target or paired supervision is used during training.

\begin{figure}[h]
  \centering
  \includegraphics[width=0.99\linewidth]{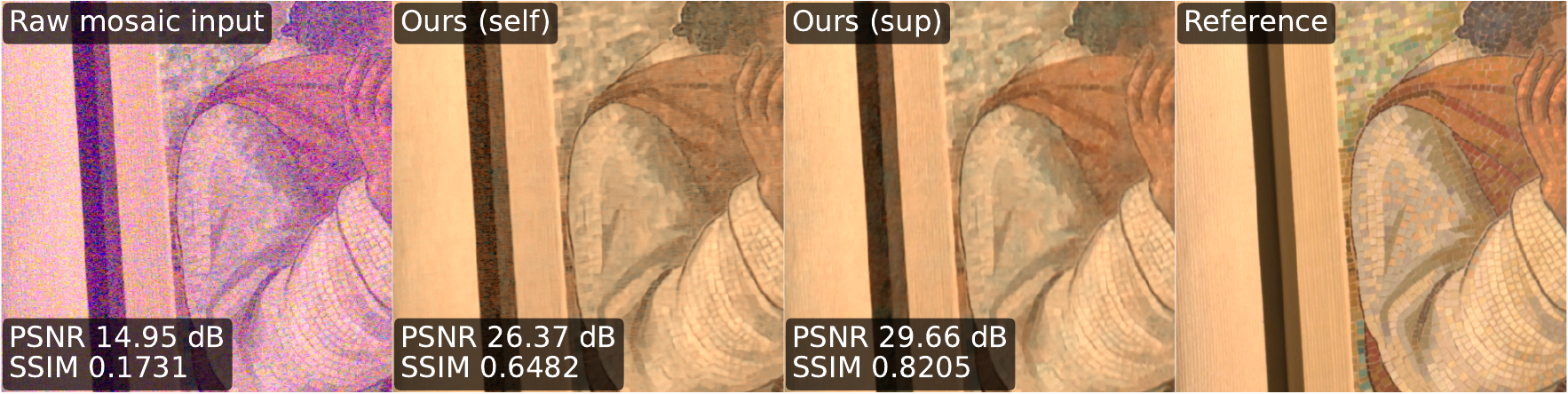}
  \caption{
  Zero-shot self-supervised reconstruction on real photon-limited data from the SID dataset. A $512\times512$ patch is reserved for testing, while the remaining patches from the same low-light image are used to train with $\mathcal{L}_{\mathrm{self}}$. The Poisson scaling parameter is estimated per RGB channel, and an affine color transform is applied for sRGB visualization and metric evaluation. The result illustrates the adaptability of the proposed method to real photon-limited sensor statistics without clean-image supervision.
  }
  \label{fig:sid_real}
\end{figure}

Since real sensor measurements can exhibit channel-dependent photon statistics, the Poisson scaling is estimated independently for each RGB channel from the raw measurements, yielding $\gamma_{\mathrm{rgb}} = (0.018,\;0.017,\;0.026).$ After reconstruction in the linear RGB domain, a standard affine color transform is applied for sRGB visualization and metric evaluation. This setting evaluates whether the proposed self-supervised reconstruction strategy can adapt to real photon-limited statistics without requiring ground-truth supervision. As shown in Fig.~\ref{fig:sid_real}, the method produces a visually stable reconstruction on real low-light data, demonstrating that the GR2R+EI training signal remains effective under realistic photon-limited acquisition conditions.

\section{Broader Scope and Future Directions}
\label{sec:supp_extensions}
The proposed framework is versatile by design: the acquisition physics are represented by a known linear forward operator, the statistical model determines the self-supervised training signal, and the image prior is provided by frozen CLIP features coupled with a lightweight trainable decoder. This separation opens two natural extension directions: adapting the self-supervised objective to broader noise models, and instantiating the same reconstruction principle for additional inverse problems.

\textbf{Extension to broader noise models.}
The current formulation uses a Poisson observation model, but the architecture itself is not tied to Poisson statistics. The frozen CLIP encoder, the lightweight decoder, and the unrolled data-consistency/prior-splitting structure can be reused with other measurement distributions. The part that changes is the statistical self-supervision module: for each noise model, the re-corruption rule and, when useful, the corresponding likelihood-based training objective should be selected according to the assumed observation distribution.

This makes the framework compatible with the broader family of noise models considered by GR2R~\cite{monroy2025gr2r}. In particular, GR2R naturally supports non-Gaussian additive noise, such as log-Rayleigh noise, as well as natural-exponential-family observation models including Gaussian, Poisson, Gamma, and Binomial noise. Therefore, extending the method to another noise distribution mainly requires replacing the Poisson-specific re-corruption procedure with the appropriate GR2R construction, while keeping the CLIP-based prior and the unrolled reconstruction architecture unchanged.

Noise-level calibration can also be incorporated within this direction. The present implementation assumes that the Poisson scaling parameter is either known from the acquisition system or estimated before training. When such calibration is unavailable, unknown-noise self-supervised objectives provide a promising alternative. For example, UNSURE and PG-UNSURE~\cite{tachella2025unsure} are designed to handle unknown noise levels by treating the noise parameters through training constraints or Lagrange multipliers. Since these objectives affect the self-supervised training signal rather than the representation prior, they are complementary to the frozen-CLIP reconstruction architecture.

\textbf{Extension to additional inverse problems.}
A second extension direction is to apply the same framework beyond demosaicing and deblurring. The original GR2R formulation for inverse problems is measurement-domain based and therefore naturally handles any linear inverse problem. This is especially attractive for computational imaging, where many acquisition systems can be modeled through a known linear operator, including inpainting, super-resolution, compressed sensing, coded-aperture imaging, accelerated MRI, sparse-view tomography, and other linear sensing pipelines~\cite{ji2026self}.

In this setting, the forward operator determines the data-consistency step, while the CLIP-based module continues to act as a reusable image prior. When the operator has convenient structure, such as diagonal masking or convolution, efficient closed-form or Fourier-domain updates can be used. For more general linear operators, the data-consistency step can be replaced by an iterative solver, such as conjugate gradients or a proximal update, without changing the learned prior. Thus, the same frozen representation backbone can be paired with different physics-driven solvers depending on the acquisition model.

For incomplete operators with non-trivial null spaces, additional self-supervised structure remains important. Equivariant imaging provides one such mechanism by exploiting transformations under which the image distribution is approximately invariant, while multi-operator acquisition settings can provide complementary measurements of the same underlying signal~\cite{tachella2026self}. These components are orthogonal to the CLIP prior and can be combined with the measurement-domain self-supervised objective to recover information not directly observed by a single forward operator.

\section{Additional Results}


This section provides additional results that complement the experiments reported in the main manuscript. In particular, Table~\ref{tab:noise_levels} reports quantitative comparisons for both Poisson demosaicing and Poisson deblurring on BSDS500 under intermediate shot-noise levels, $\gamma=0.02$ and $\gamma=0.03$.  Extended visual comparisons for both Poisson demosaicing and Poisson deblurring under different shot-noise levels and across BSDS500 and DIV2K are also included. These qualitative results complement the quantitative evaluation by illustrating the behavior of the proposed method in both in-distribution and cross-dataset settings, and under both mild and severe photon-limited conditions. For each experiment, reconstructions from the competing methods are shown together with the self-supervised and supervised versions of the proposed approach.

\begin{table}[h]
\centering
\caption{Additional BSDS500 results at intermediate Poisson noise levels.}
\label{tab:noise_levels}
\small
\setlength{\tabcolsep}{4pt}
\renewcommand{\arraystretch}{1.10}
\resizebox{\linewidth}{!}{%
\begin{tabular}{l c cc|cc}
\hline
\multirow{3}{*}{\textbf{Method}} &  & \multicolumn{4}{c}{\textbf{Poisson Noise}} \\
\cline{3-6}
& \multirow{2}{*}{\textbf{$\gamma$}} &
\multicolumn{2}{c|}{\textbf{Demosaicing}} & \multicolumn{2}{c}{\textbf{Deblurring}} \\
\cline{3-6}
& & \textbf{PSNR [dB]} & \textbf{SSIM} & \textbf{PSNR [dB]} & \textbf{SSIM} \\
\hline
DPIR            & \multirow{6}{*}{\textbf{0.02}} & 26.75$\pm$1.83 & 0.7463$\pm$0.0813 & 27.89$\pm$2.17 & 0.7738$\pm$0.0684 \\
Transfer CLIP   &                                  & 24.53$\pm$2.91 & 0.6608$\pm$0.1315 & 22.96$\pm$2.33 & 0.6909$\pm$0.0700 \\
GSPnP           &                                  & 28.31$\pm$1.76 & 0.7798$\pm$0.0726 & 28.44$\pm$2.22 & 0.7962$\pm$0.0690 \\
RAM  &                                  & 28.73$\pm$1.94 & 0.8084$\pm$0.0041 & 27.39$\pm$2.45 & 0.7540$\pm$0.0728 \\
\textbf{Ours (Self)} &                              & \textbf{29.00$\pm$1.77} & \textbf{0.8284$\pm$0.0511} & \textbf{28.69$\pm$2.24} & \textbf{0.8166$\pm$0.0614} \\
\textbf{Ours (Sup)}  &                              & \textbf{29.14$\pm$1.76} & \textbf{0.8368$\pm$0.0480} & \textbf{28.85$\pm$2.26} & \textbf{0.8238$\pm$0.0614} \\
\hline
DPIR            & \multirow{6}{*}{\textbf{0.03}} & 26.24$\pm$1.78 & 0.7182$\pm$0.0784 & 27.36$\pm$2.13 & 0.7547$\pm$0.0733 \\
Transfer CLIP   &                                  & 22.35$\pm$2.64 & 0.5663$\pm$0.1379 & 22.03$\pm$2.15 & 0.6367$\pm$0.0756 \\
GSPnP           &                                  & 27.44$\pm$1.81 & 0.7491$\pm$0.0831 & 27.82$\pm$2.19 & 0.7736$\pm$0.0753 \\
RAM  &                                  & 27.63$\pm$1.80 & 0.7699$\pm$0.0045 & 26.72$\pm$2.35 & 0.7236$\pm$0.0742 \\
\textbf{Ours (Self)} &                              & \textbf{28.16$\pm$1.71} & \textbf{0.8030$\pm$0.0547} & \textbf{28.19$\pm$2.19} & \textbf{0.7998$\pm$0.0658} \\
\textbf{Ours (Sup)}  &                              & \textbf{28.38$\pm$1.76} & \textbf{0.8132$\pm$0.0544} & \textbf{28.36$\pm$2.22} & \textbf{0.8070$\pm$0.0668} \\
\hline
\end{tabular}%
}
\end{table}

\subsection{Additional Poisson Noise Levels}
\label{sec:supp_additional_noise_levels}

Table~\ref{tab:noise_levels} reports additional BSDS500 experiments at intermediate Poisson noise levels, $\gamma=0.02$ and $\gamma=0.03$, for both CFA demosaicing and deblurring. These settings complement the noise regimes evaluated in the main experiments and provide a finer view of the reconstruction behavior as the photon-limited degradation increases.

The results are consistent across tasks and noise levels. As expected, performance decreases smoothly from $\gamma=0.02$ to $\gamma=0.03$, while the proposed self-supervised variant remains competitive and consistently outperforms the evaluated baselines. For demosaicing, Ours (Self) improves over the strongest competing baseline by $+0.27$ dB at $\gamma=0.02$ and $+0.53$ dB at $\gamma=0.03$. For deblurring, the corresponding gains are $+0.25$ dB and $+0.37$ dB. The supervised variant provides only a small upper bound over the self-supervised model, with gaps of $0.14$--$0.22$ dB for demosaicing and $0.16$--$0.17$ dB for deblurring. These trends indicate that the proposed GR2R+EI training objective remains stable across intermediate photon-limited regimes and that the observed performance is not tied to a small set of selected noise levels.

\subsection{Additional Qualitative Results}
\label{sec:supp_additional_noise_levels}

\begin{figure}[h]
    \centering
    \includegraphics[width=\linewidth]{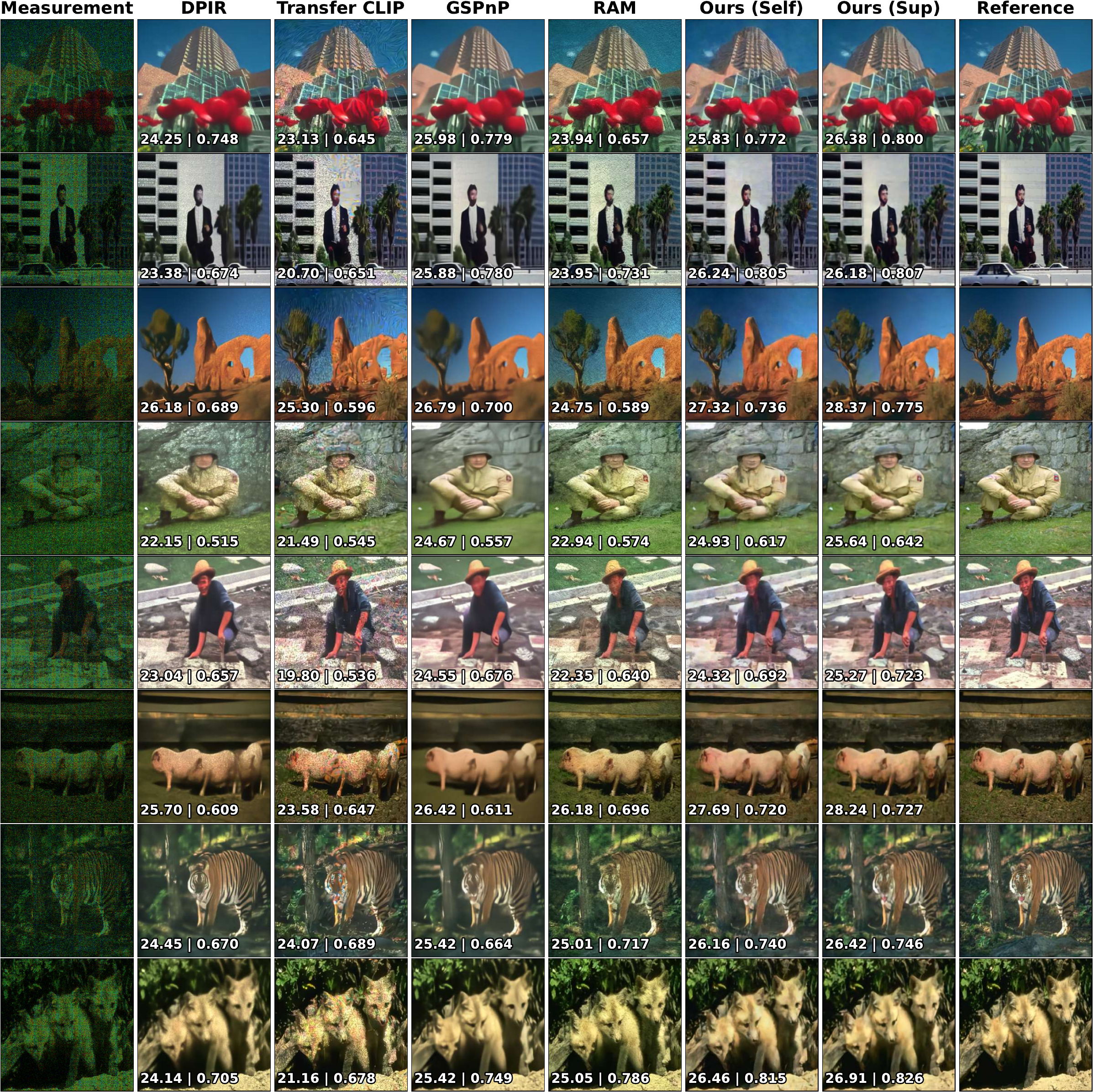} 
    \caption{Qualitative comparison on BSDS500 for Poisson demosaicing with shot-noise level $\gamma=0.05$. Columns show the mosaiced noisy \emph{Measurement}, reconstructions by DPIR, Transfer CLIP, GSPnP, and RAM, followed by the proposed method trained self-supervised (Ours~(Self)) and its supervised counterpart (Ours~(Sup)), and the \emph{Reference}. Per-image PSNR/SSIM are overlaid for direct visual comparison.}
    \label{fig:demosaicing005bsds500}
\end{figure}

\begin{figure}[h]
    \centering
    \includegraphics[width=\linewidth]{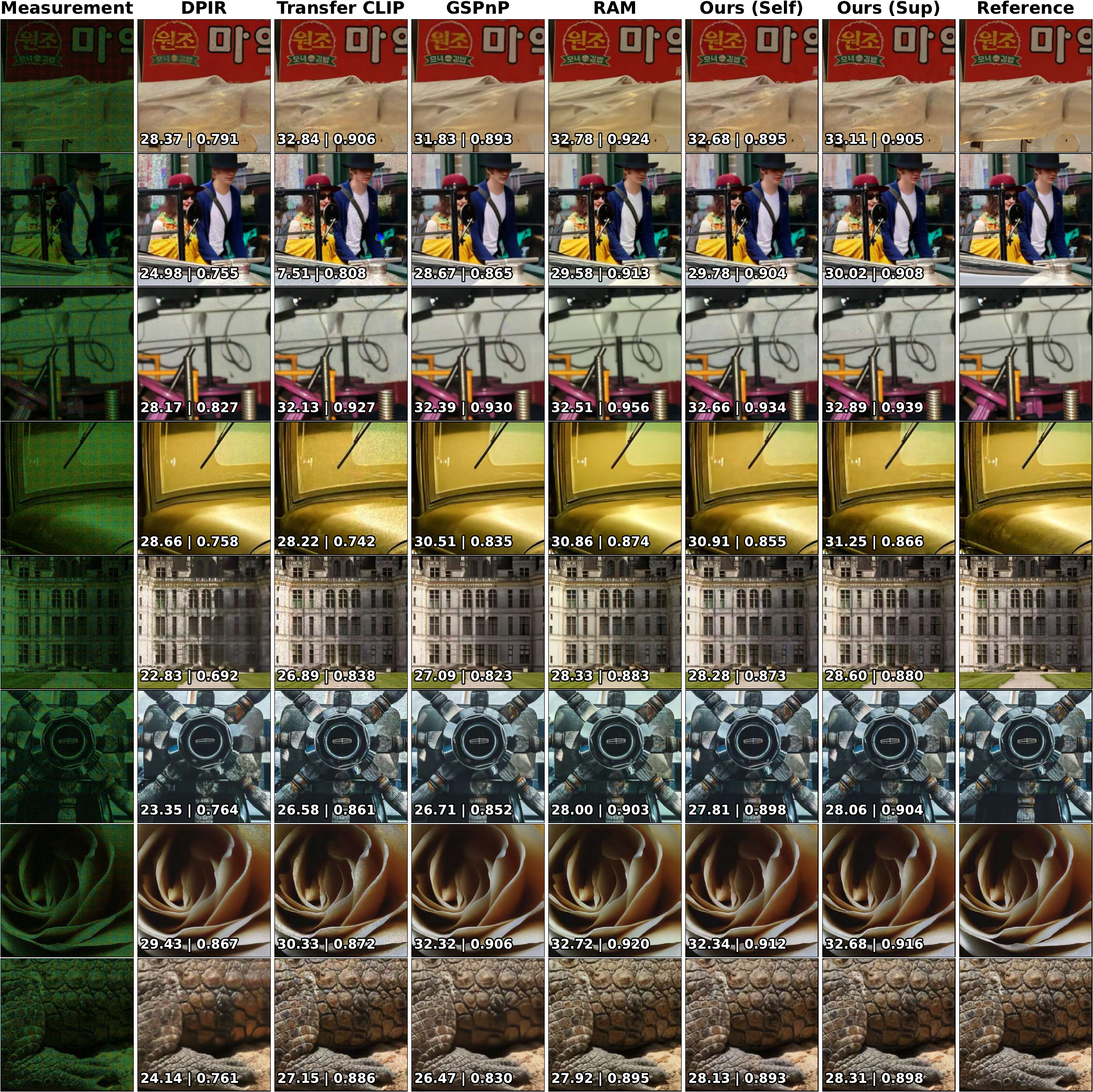} 
    \caption{
Qualitative comparison on DIV2K for Poisson demosaicing with shot-noise level $\gamma=0.01$.
Columns show the mosaiced noisy \emph{Measurement}, reconstructions by DPIR, Transfer CLIP, GSPnP, and RAM, followed by the proposed method trained self-supervised (Ours~(Self)) and its supervised counterpart (Ours~(Sup)), and the \emph{Reference}.
Per-image PSNR/SSIM are overlaid for direct visual comparison.
}
    \label{fig:demosaicing001div2k}
\end{figure}

\begin{figure}[h]
    \centering
    \includegraphics[width=\linewidth]{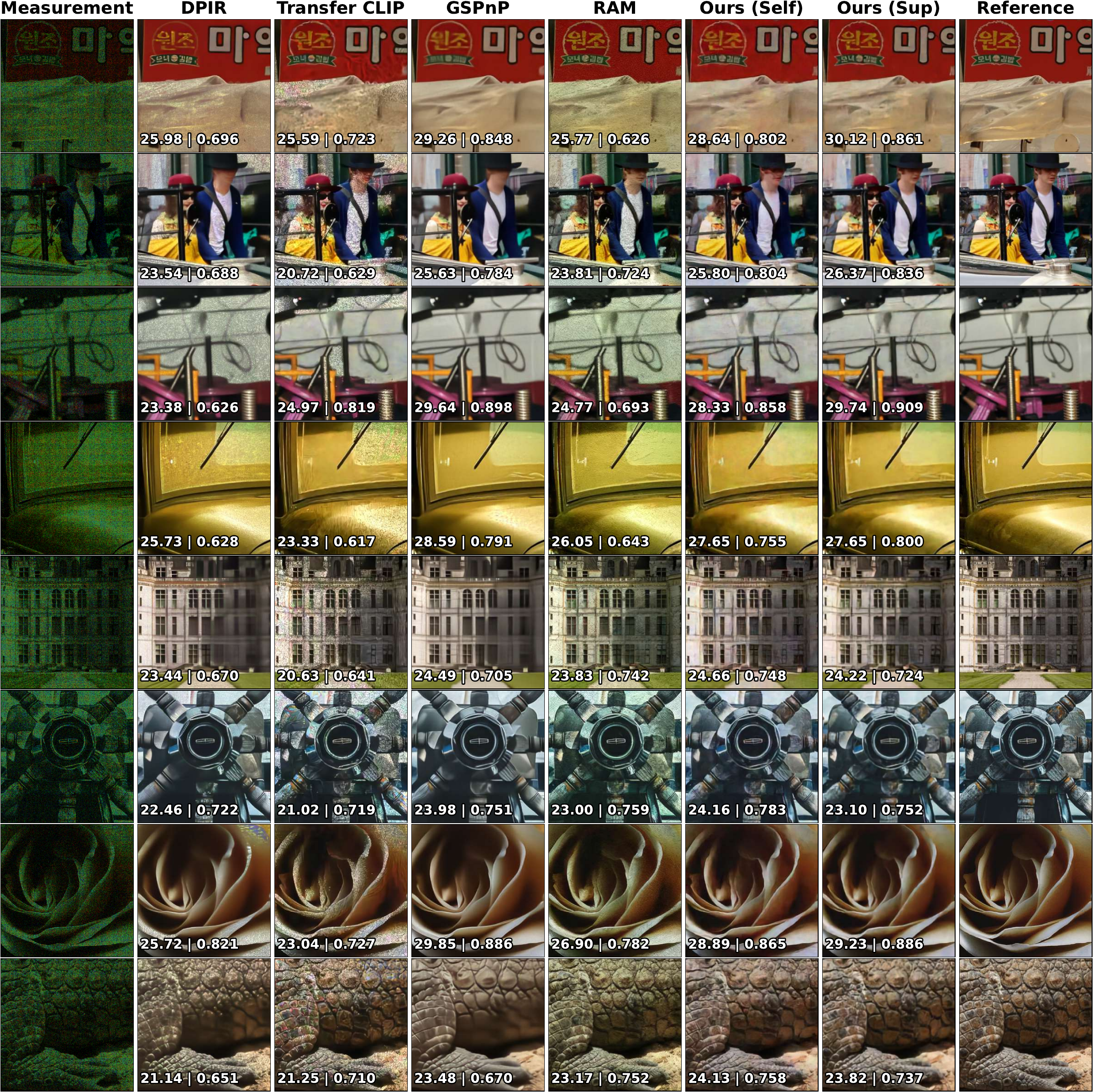} 
    \caption{
Qualitative comparison on DIV2K for Poisson demosaicing with shot-noise level $\gamma=0.05$.
Columns show the mosaiced noisy \emph{Measurement}, reconstructions by DPIR, Transfer CLIP, GSPnP, and RAM, followed by the proposed method trained self-supervised (Ours~(Self)) and its supervised counterpart (Ours~(Sup)), and the \emph{Reference}.
Per-image PSNR/SSIM are overlaid for direct visual comparison.
}
    \label{fig:demosaicing005div2k}
\end{figure}

\begin{figure}[h]
    \centering
    \includegraphics[width=\linewidth]{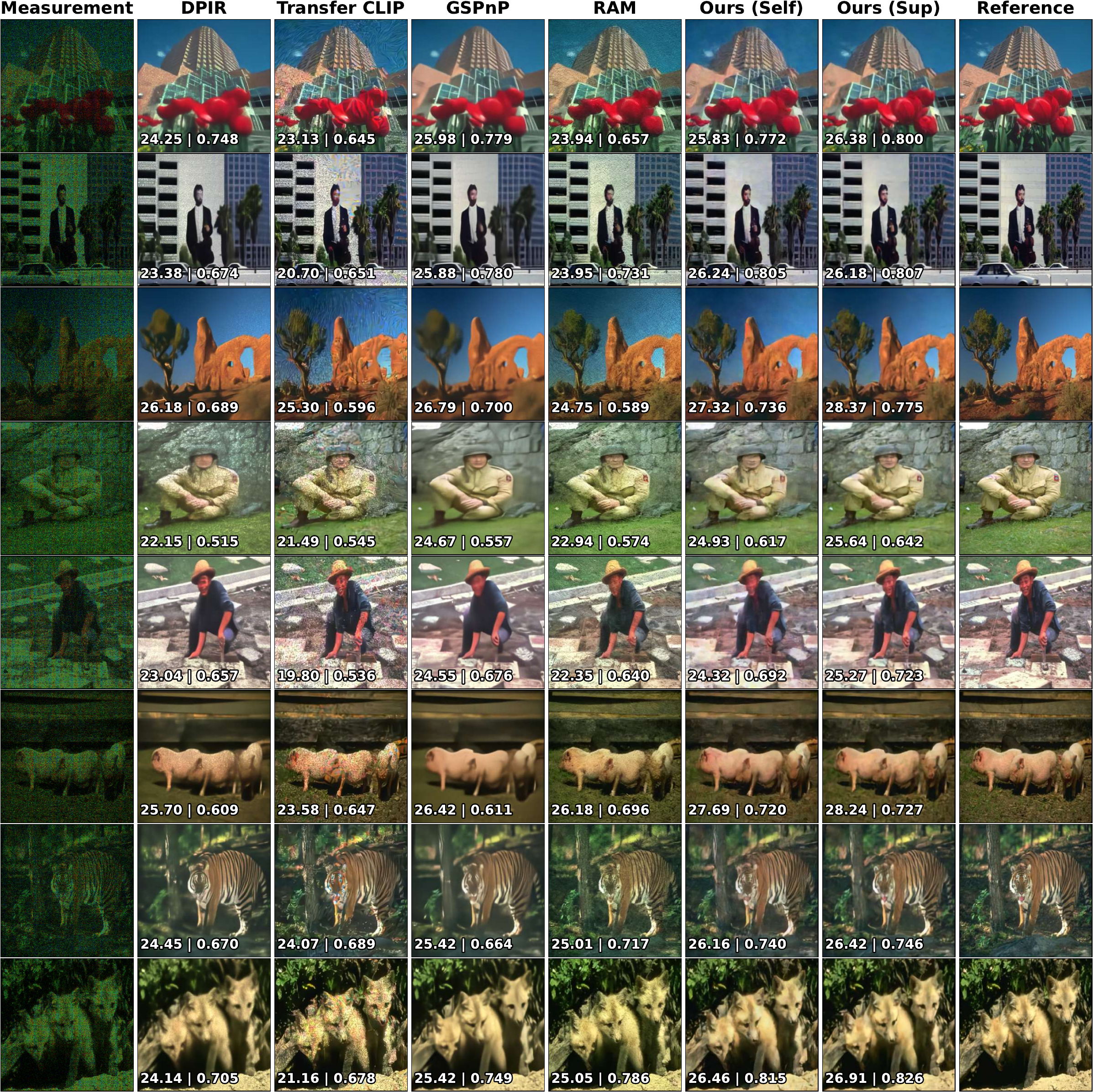} 
    \caption{
Qualitative comparison on BSDS500 for Poisson deblurring with shot-noise level $\gamma=0.05$.
Columns show the blurred noisy \emph{Measurement}, reconstructions by DPIR, Transfer CLIP, GSPnP, and RAM, followed by the proposed method trained self-supervised (Ours~(Self)) and its supervised counterpart (Ours~(Sup)), and the \emph{Reference}.
Per-image PSNR/SSIM are overlaid for direct visual comparison.
}
    \label{fig:deblurring005bsds500}
\end{figure}

\begin{figure}[h]
    \centering
    \includegraphics[width=\linewidth]{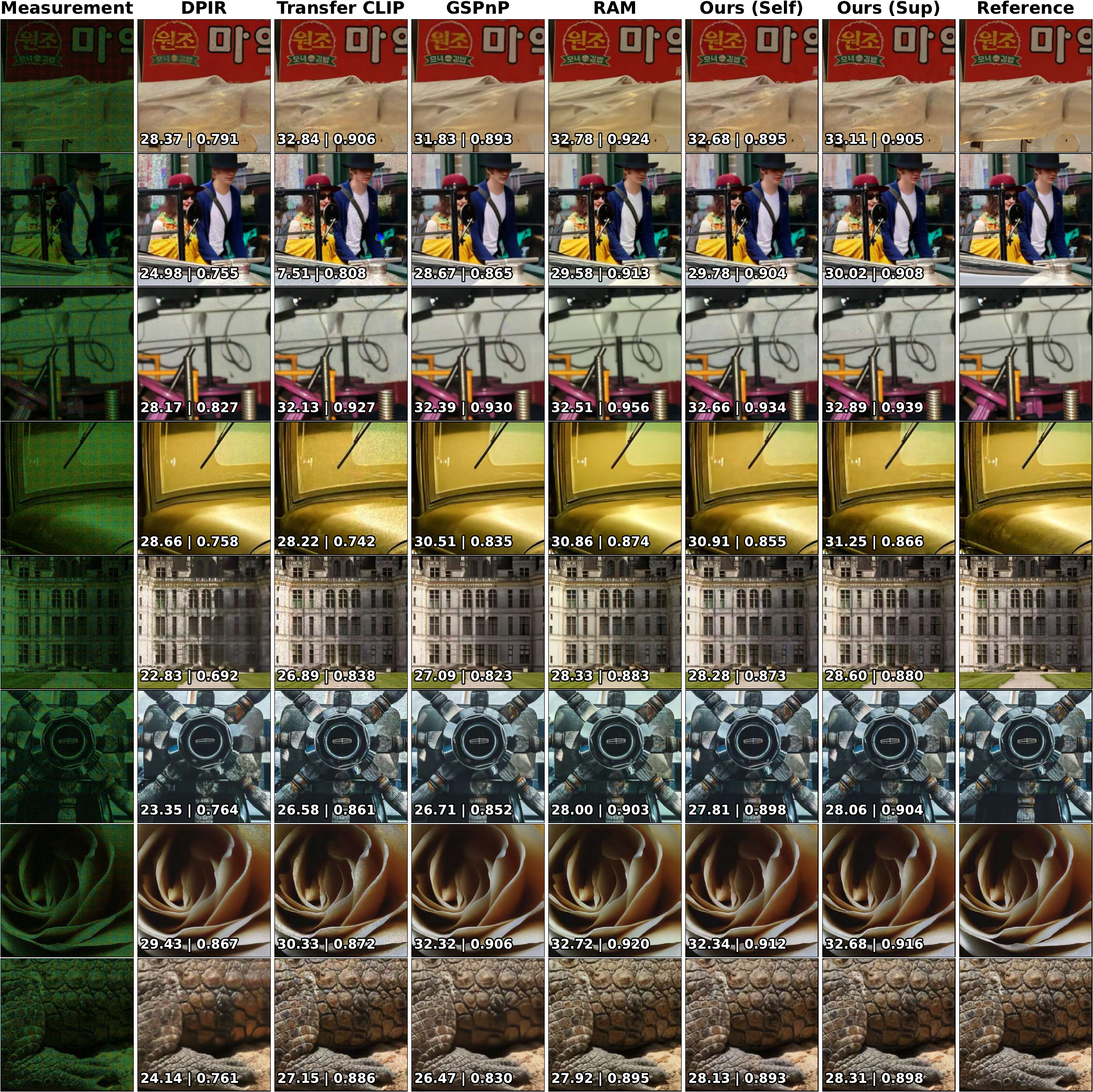} 
    \caption{
Qualitative comparison on DIV2K for Poisson deblurring with shot-noise level $\gamma=0.01$.
Columns show the blurred noisy \emph{Measurement}, reconstructions by DPIR, Transfer CLIP, GSPnP, and RAM, followed by the proposed method trained self-supervised (Ours~(Self)) and its supervised counterpart (Ours~(Sup)), and the \emph{Reference}.
Per-image PSNR/SSIM are overlaid for direct visual comparison.
}
    \label{fig:deblurring001div2k}
\end{figure}

\begin{figure}[h]
    \centering
    \includegraphics[width=\linewidth]{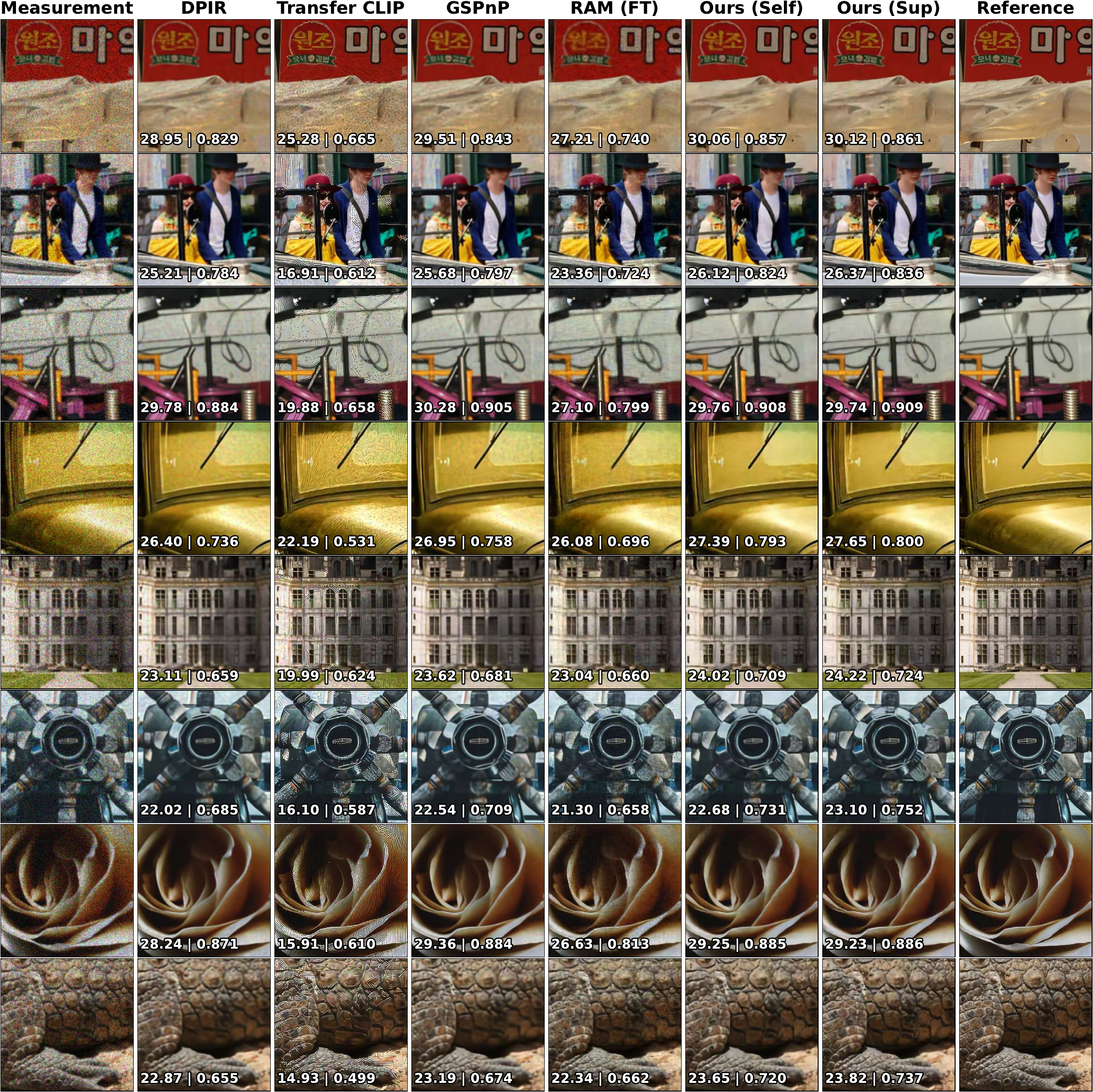} 
    \caption{
Qualitative comparison on DIV2K for Poisson deblurring with shot-noise level $\gamma=0.05$.
Columns show the blurred noisy \emph{Measurement}, reconstructions by DPIR, Transfer CLIP, GSPnP, and RAM, followed by the proposed method trained self-supervised (Ours~(Self)) and its supervised counterpart (Ours~(Sup)), and the \emph{Reference}.
Per-image PSNR/SSIM are overlaid for direct visual comparison.
}
    \label{fig:deblurring005div2k}
\end{figure}

Figures~\ref{fig:demosaicing005bsds500}, \ref{fig:demosaicing001div2k}, and~\ref{fig:demosaicing005div2k} show additional qualitative results for Poisson demosaicing. These examples cover the challenging BSDS500 setting at $\gamma=0.05$ and the DIV2K dataset-shift setting at $\gamma=0.01$ and $\gamma=0.05$. Figures~\ref{fig:deblurring005bsds500}, \ref{fig:deblurring001div2k}, and~\ref{fig:deblurring005div2k} provide the corresponding qualitative results for Poisson deblurring. As in the demosaicing case, we report results on BSDS500 at $\gamma=0.05$ and on DIV2K at both $\gamma=0.01$ and $\gamma=0.05$. Each figure compares the noisy \emph{Measurement}, the reconstructions obtained by the competing methods, the proposed self-supervised and supervised variants, and the clean \emph{Reference} image.

\newpage

\noindent\textbf{Poisson Demosaicing on BSDS500 at $\gamma=0.05$}

\newpage

\noindent\textbf{Poisson Demosaicing on DIV2K at $\gamma=0.01$}

\newpage

\noindent\textbf{Poisson Demosaicing on DIV2K at $\gamma=0.05$}

\newpage

\noindent\textbf{Poisson Deblurring on BSDS500 at $\gamma=0.05$}

\newpage

\noindent\textbf{Poisson Deblurring on DIV2K at $\gamma=0.01$}

\newpage

\noindent\textbf{Poisson Deblurring on DIV2K at $\gamma=0.05$}

\newpage


\bibliographystyle{splncs04}
\bibliography{main}

\end{document}